\documentclass[runningheads,fleqn]{cl2emult}

\usepackage{makeidx}  
\usepackage{graphicx} 
\usepackage{subeqnar} 
\usepackage{multicol} 
\usepackage{cropmark} 
\usepackage{eso}      
\makeindex            

\begin{document}
\title*{A Sunyaev-Zel'dovich Effect Survey for 
High Redshift Clusters}
\toctitle{A Sunyaev-Zel'dovich Effect Survey for 
\protect\newline High Redshift Clusters}
%
%
\titlerunning{SZE Survey for High-z Clusters}
%
\author{J.J. Mohr\inst{1,4}
\and J.E. Carlstrom\inst{1}
\and G.P. Holder\inst{1}
\and W.L. Holzapfel\inst{2}
\and M.K. Joy\inst{3}
\and E.M. Leitch\inst{1}
\and E.D. Reese\inst{1}}
\authorrunning{Mohr et al.}
%
%
\institute{University of Chicago, Chicago, IL USA
\and University of California at Berkeley, Berkeley, CA  USA
\and NASA/Marshall Space Flight Center, Huntsville, AL,  USA}

\maketitle              

\medskip
\centerline{\small\it{to appear in the proceedings of the VLT Opening 
Symposium}}

\begin{abstract}
 
Interferometric observations of the Sunyaev-Zel'dovich Effect (SZE) toward
clusters of galaxies provide sensitive cosmological probes.  We present
results from 1~cm observations (at BIMA and OVRO) of a large, intermediate
redshift cluster sample.  In addition, we describe a proposed, higher
sensitivity array which will enable us to survey large portions of 
the sky. Simulated observations indicate that we will be able to survey one
square degree of sky per month to sufficient depth that we will detect all
galaxy clusters more massive than $2\times10^{14} h^{-1}_{50}\,
M_\odot$, regardless of their redshift.  We describe the cluster yield and 
resulting cosmological constraints from such a survey.

\end{abstract}

\section{Interferometric Sunyaev-Zel'dovich Effect Imaging}
\footnotetext[4]{Chandra Fellow}
Interactions between cosmic microwave background (CMB) photons and the hot, 
ionized plasma in galaxy clusters introduce changes in the spectrum 
of the CMB.  At wavelengths of 1~cm we observe 
this Sunyaev-Zel'dovich Effect (SZE) 
\cite{sunyaev72} as a decrease in the intensity of the 
CMB; this decrement $\Delta T(R)$ at projected cluster radius $R$ can be 
expressed as
\begin{equation}
{\Delta T(R)\over T_{CMB}}= -2{\sigma_{T}\over m_{e}c^{2}}\int dl\, 
n_{e}(l,R) k_{B}T_{e}(l,R)
\label{eq:deltaT}
\end{equation}
where $\sigma_{T}$ is the Thomson cross section, $m_{e}c^{2}$ is the 
electron rest mass, $n_{e}$ is the electron number density and 
$k_{B}$ is the Boltzmann constant.  Note that the fractional change in 
the CMB temperature $\Delta T/T_{CMB}$ is independent of redshift; it 
depends only on properties of the ICM.  Higher order corrections to 
this expression and a consideration of the effects of cluster peculiar 
velocities on the CMB spectrum can be found elsewhere 
(e.g., \cite{birkinshaw99}).

The typical central decrement for a massive $\sim10^{15}\ M_{\odot}$ 
galaxy cluster is $\sim1$~mK, and for typical radio telescope 
system temperatures the SZE decrement is only a few parts in $10^{5}$ of 
the power detected by the receiver, making SZE observations 
challenging.  Characteristics of radio interferometry make it
well suited to meet this challenge.  Perhaps the most important of these is
that only the correlated portion of the receiver output is signal, while 
the uncorrelated output serves as noise; differential ground 
spillover and detector noise are naturally separated from the cluster 
decrement.  Another is that interferometry effectively simultaneously
differences over the sky, easily filtering the smoothly distributed 
$\sim$3~K background from the cluster SZE decrement we seek.
Finally, with interferometry one obtains information on many scales
simultaneously; by placing some dishes at larger separations it is 
straightforward to filter out point source contributions.

\begin{figure}[htb]
\vskip-10pt
\includegraphics[width=1.0\textwidth]{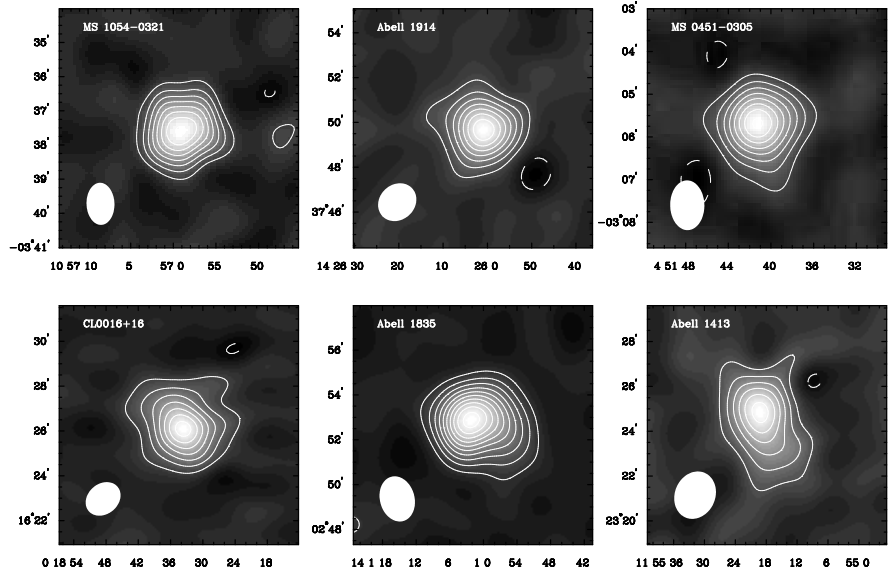}
\caption[]{SZE decrement images of six of the 27 clusters observed with the 
Sunyaev-Zel'dovich Imaging Experiment at BIMA and OVRO.  Contours are 
spaced by 2$\sigma$, and the synthesized beam is represented by the 
white ellipse in the lower left of each frame.
\label{fig:szclusters}}
\vskip-10pt
\end{figure}

Over the past several years, Carlstrom, Joy and collaborators have 
developed sensitive, low noise receivers and used them at the
BIMA and OVRO observatories to obtain high quality SZE data 
on a sample of 27 intermediate redshift clusters \cite{carlstrom96}.  Figure 
\ref{fig:szclusters} contains greyscale SZE images and signal to noise 
contour overlays of 6 clusters. The contours are spaced by 2$\sigma$, 
and the white ellipses in the lower left of each frame
represent the FWHM of the synthesized beam.

As demonstrated by Figure \ref{fig:szclusters}, the interferometric SZE 
Imaging Experiment does deliver 
high signal to noise SZE images of intermediate redshift 
clusters.  We use these SZE data to constrain the cluster ICM
distribution, and, in combination with published measurements of the 
ICM temperature, these data enable estimates of cluster ICM masses and 
mass fractions \cite{grego99a,grego99b}.
In combination with X-ray images of galaxy clusters, the SZE data 
allow one to estimate cluster sizes and measure direct, 
distance-ladder-independent distances.  The X-ray surface brightness 
$I(R)$ at projected cluster radius $R$ can be expressed as
\begin{equation}
I_{X}(R)={1\over4\pi (1+z)^{4}}{\mu_{e}\over\mu_{H}}\int dl\, 
n^{2}_{e}(l,R) \Lambda(T_{e})
\label{eq:xray}
\end{equation}
where $n_{e}$ is the electron number density, $\Lambda(T_{e})$ is the 
temperature dependent cooling coefficient and 
$n_{e}=\rho/m_{p}\mu_{e}$.  Note that $I(R)$ depends on the square of 
the electron density, and is highly redshift dependent.  Because of 
the different dependence on the electron density, observations of 
$I(R)$ and $\Delta T/T_{CMB}$ provide a direct measure of the line of 
sight distance through the cluster.  With additional assumptions 
about the cluster geometry, we can combine this line of sight size 
with the angular size of the cluster on the sky to estimate 
the cluster angular diameter distance $D_{A}$.  We are currently using 
archival X-ray data to finalize distance estimates in our ensemble of 
27 SZE clusters \cite{reese99}.

\begin{figure}[htb]
\vskip-10pt
\hbox to \hsize{
\includegraphics[width=0.55\textwidth]{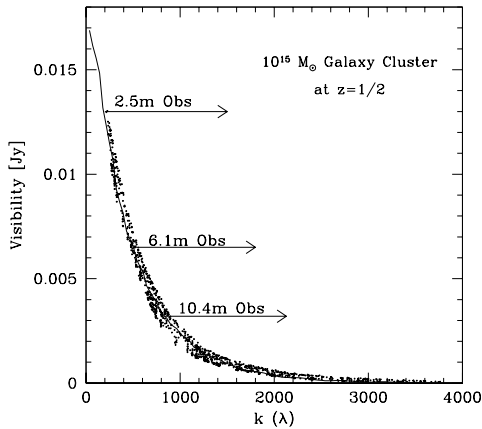}\hfill
\vbox to 2.6in{\hsize=2.0in
\caption[]{One advantage of 2.5~m telescopes over the currently 
available 6.1~m BIMA and 10~m OVRO dishes is clear in this plot of 
cluster visibility versus $k$ which is the telescope 
separation in units of the observing wavelength (1~cm).  The 
cluster visibility falls off exponentially with $k$, yielding significant 
gains for smaller telescopes which can sample the visibility at 
smaller separations.  The points are visibilities from a simulated 
galaxy cluster, and the line is the best fit model.
\label{fig:dishes}}\vfill}}
\vskip-10pt
\end{figure}

\section{The Proposed SZE Interferometric Array}

The difficulty of detecting X-ray emission from high redshift 
clusters-- 
even with the high quality X-ray instruments available today-- is made 
clear by Eqn. \ref{eq:xray}.
Indeed, it is the redshift independent nature of the SZE (see Eqn. 
\ref{eq:deltaT}) that makes it an ideal tool for studying collapsed 
structures at high redshift.  The current best instruments for SZE 
studies of high redshift clusters do not have the sensitivity required 
to detect sizeable numbers of clusters in a non-targeted search within 
reasonable observing times of 1~mth to 1~yr.  Therefore, we have 
proposed to build a new array composed of ten 2.5~m dishes and a 
wide-band digital correlator.  This array, when operating at 1~cm, is 
optimized for high redshift cluster 
detection and will be approximately 100 times more sensitive than 
our current SZE Imaging Experiment on BIMA.

In this era of large telescopes it may be surprising to learn that 
the best telescope size for SZE cluster studies is 2.5~m.  There are two 
main reasons for this.  First, the angular size $\theta_{BEAM}$ of the primary 
beam of a radio telescope scales with the dish diameter $D$ as 
$\theta_{BEAM}\propto\lambda_{obs}/D$, where $\lambda_{obs}$ is the 
observation wavelength.  $\theta_{BEAM}$ determines the field 
size in a survey, and cluster signal on scales larger than
$\theta_{BEAM}/2$ is lost.  For 1~cm observations on a 6.1~m dish, 
the FWHM of the primary beam is $\theta_{BEAM}\sim6$~arcmin, whereas 
for the 2.5~m telescopes $\theta_{BEAM}\sim15$~arcmin.  The primary 
beam of a 2.5~m dish is better matched to cluster angular scales than 
a 6.1~m dish.  Second, as is shown graphically in Figure 
\ref{fig:dishes}, interferometry samples the cluster visibility, 
which is the Fourier transform of the CMB decrement distribution on the 
sky.  Cluster visibilities typically fall off exponentially with 
distance from the origin $k$ (as opposed to point sources whose Fourier 
transforms are independent of $k$).  Therefore, with smaller dishes 
one samples a region of the U-V plane where there is significantly 
more cluster flux.

\begin{figure}[htb]
\vskip-10pt
\hbox to \hsize{
\includegraphics[width=0.6\textwidth]{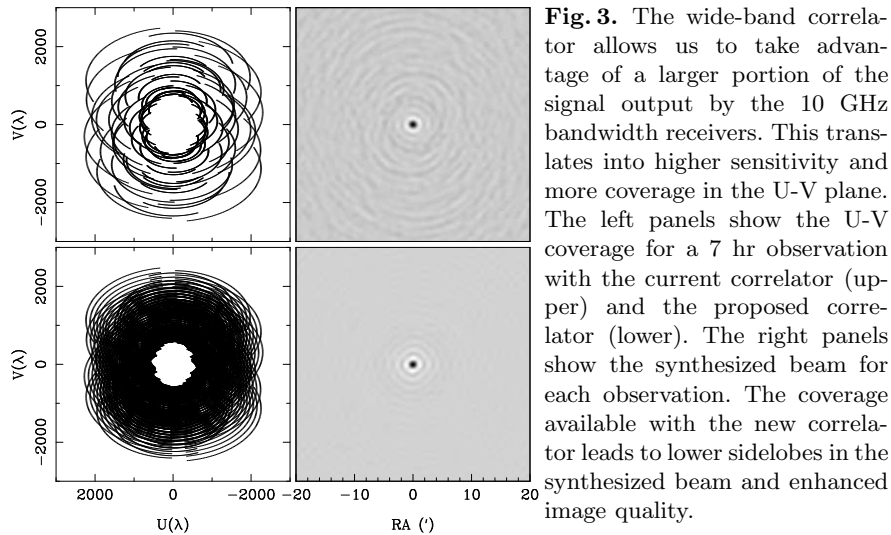}
\hfill\vbox to 2.85in{\hsize=1.8in
\caption[]{The wide-band correlator allows us to take advantage of a 
larger portion of the signal output by the 10~GHz bandwidth 
receivers.  This translates into higher sensitivity and more 
coverage in the U-V plane.  The left panels show the U-V coverage for 
a 7~hr observation with the current correlator (upper) and the 
proposed correlator (lower).  The right panels show the synthesized 
beam for each observation.  The coverage available with 
the new correlator leads to lower sidelobes in the synthesized beam and 
enhanced image quality.
\label{fig:wideband}}\vfil}}
\vskip-10pt
\end{figure}

The proposed wide-band digital correlator will enable us to use a 
larger fraction of the signal output from the 10~GHz bandwidth receivers.  
Currently, the maximum bandwidth at BIMA is $\sim$800~MHz; thus, 
the proposed 8~GHz correlator will increase the sensitivity by an order of 
magnitude.  In addition to higher throughput, the wide-band correlator 
provides enhanced image quality.  As is apparent in Figure 
\ref{fig:wideband}, the fraction of the U-V plane sampled in 
a 7~hr exposure with the old correlator (top left) is much smaller than 
the fraction sampled in the same exposure with the new correlator (bottom left). 
This additional U-V coverage leads to a synthesized beam (right 
panels) which has smaller sidelobes.  This enhanced image quality 
simplifies structure detection in SZE images.

\section{The SZE Galaxy Cluster Survey}

X-ray samples of high redshift clusters are 
disappointingly small, primarily because of the strong redshift 
dependence of cluster emission (see Eqn. \ref{eq:xray}); only the most luminous 
clusters at each redshift tend to be found.  However, with this 
sensitive new SZE array, we will be able to carry out a survey of 
$\sim1$~deg$^{2}$ per month and detect all clusters more massive than
$2\times10^{14}h^{-1}_{50}$~$M_{\odot}$, independent of their redshift.
Interestingly, the cluster selection function from such a survey is 
essentially a cut in cluster binding mass.  So not only will an SZE 
survey yield a sizeable high redshift cluster sample for the first 
time, but the survey will also deliver a sample which is much easier to 
understand than typical X-ray samples at intermediate redshift.

\begin{figure}[htb]
\vskip-10pt
\hbox to \hsize{
\includegraphics[width=0.48\textwidth]{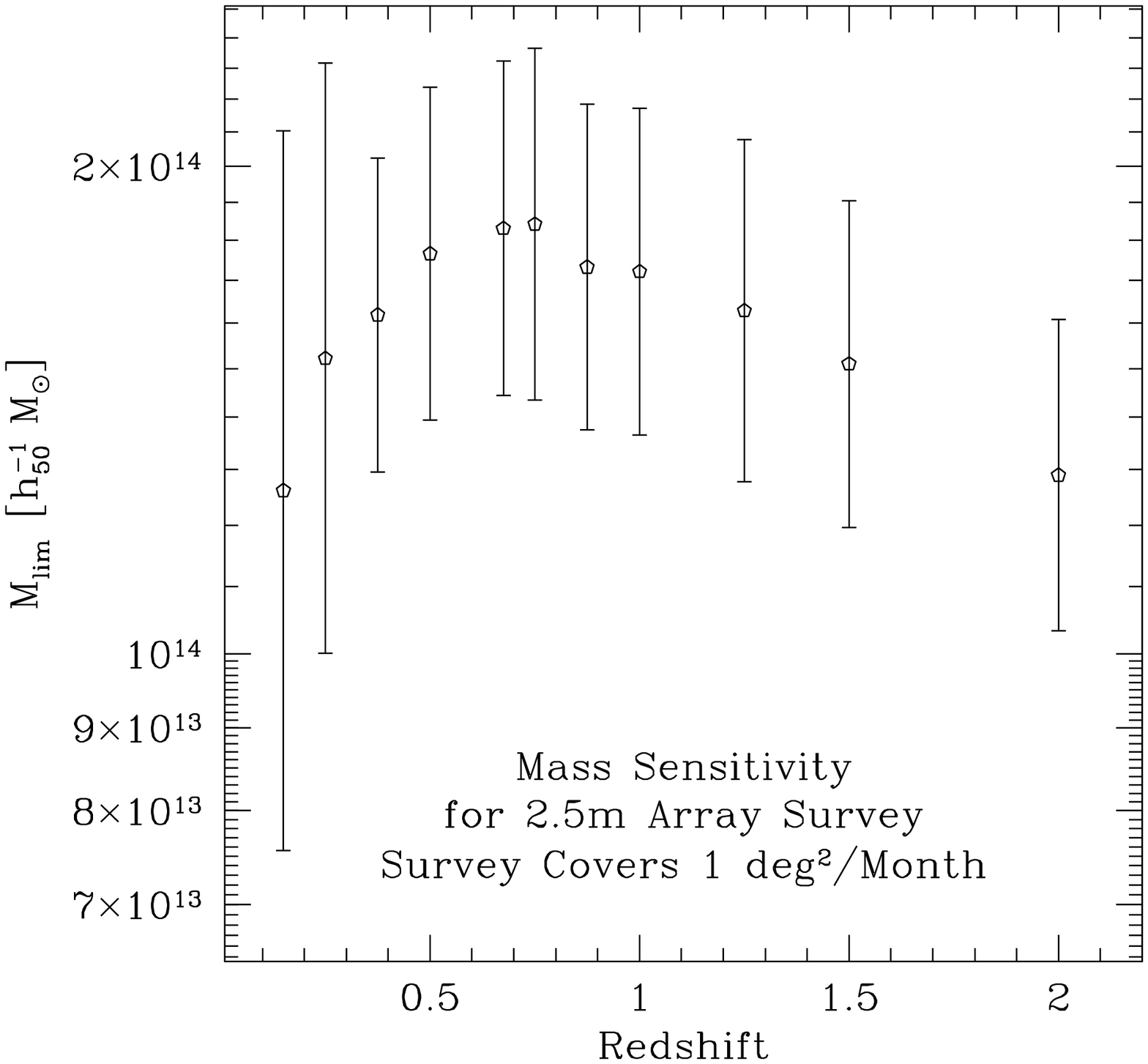}
\includegraphics[width=0.48\textwidth]{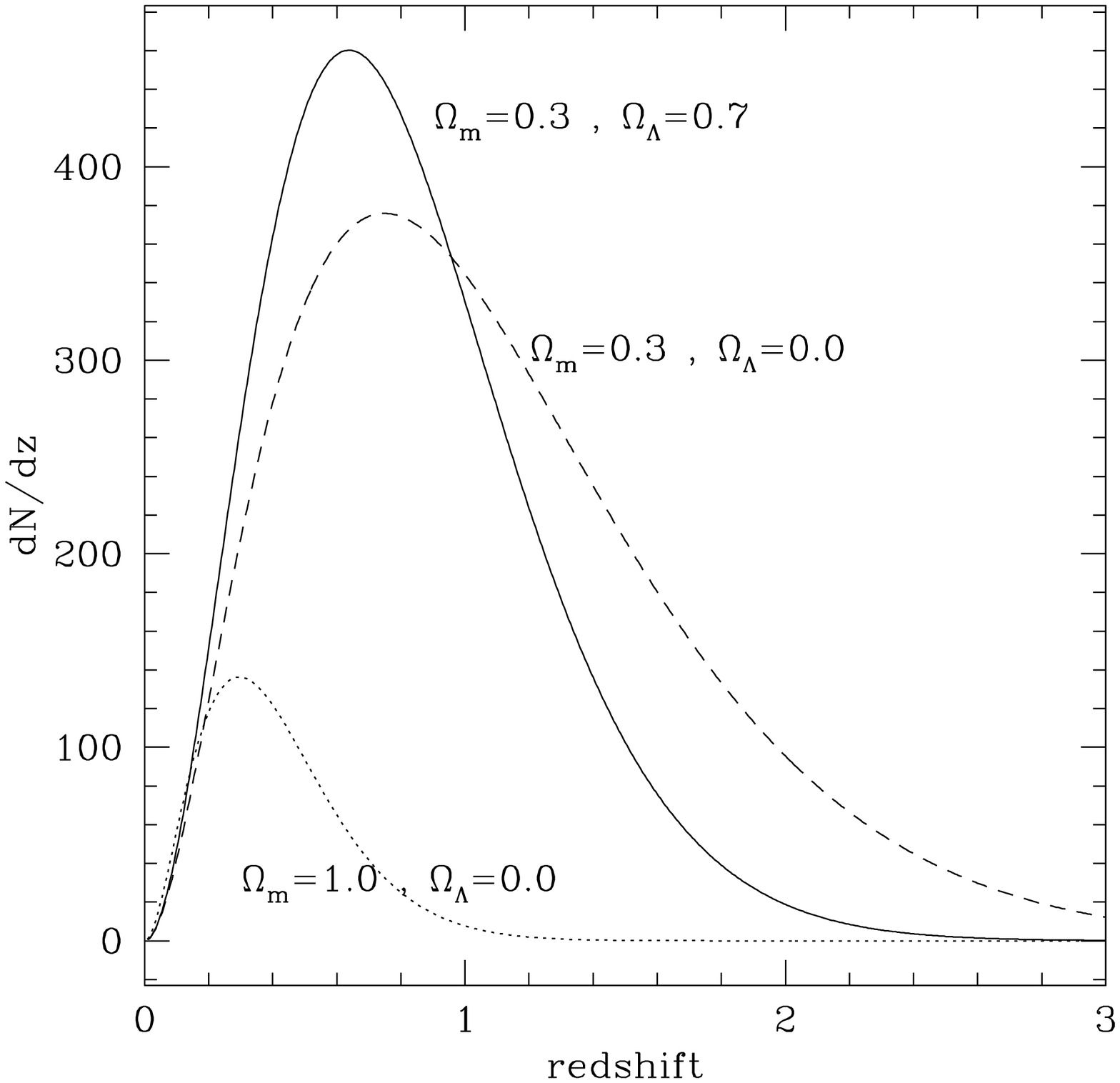}}
\caption[]{
The limiting mass $M_{lim}$
of our proposed SZE survey 
versus redshift (left).  We expect to detect all clusters more 
massive than $2\times10^{14}h^{-1}_{50}$~$M_{\odot}$, independent of 
their redshift!  To the right we plot the abundance of clusters
more massive than $M_{lim}$ as a function of redshift.  These 
Press-Schechter models are normalized to produce the observed 
cluster abundance in the nearby universe.
\label{fig:selection}}
\vskip-10pt
\end{figure}

We determine the survey mass limit as a function of redshift through mock 
observations of N-body plus SPH simulations of galaxy clusters.  These 
mock SZE observations include noise and have exposures tuned so that 
we will be able to survey 1~deg$^{2}$ in a month.  The N-body plus 
SPH simulations are state of the art simulations sampled from four 
different cosmological models: (1) SCDM: standard, biased 
$\Omega=1$ CDM, (2) OCDM: unbiased $\Omega_{m}=0.3$ CDM, (3) LCDM: 
unbiased $\Omega_{m}=0.3$ and $\Omega_{\lambda}=0.7$ CDM, and (4) 
$\tau$CDM: $\Omega_{m}=1$ but CDM power spectrum with $\Gamma=0.24$
(see \cite{mohr97,mohr99} for more details).  These simulations 
include only gas dynamics and gravity.  We use three particle species; 
this allows us to follow tidal fields from surrounding large scale 
structure while still resolving the dynamics in the cluster core.  
We incorporate no early heating of the ICM from galaxy formation.

To determine the detection limit at each redshift we use outputs of 
the 48 simulations from the appropriate redshift; with this approach the 
morphological evolution of galaxy clusters is naturally included.  We 
determine the detection probability in each mock observation by 
fitting a cluster model and determining the 
$\Delta\chi^{2}$ between the best fitting and null models.  
We set our detection threshold at 5$\sigma$.

Figure \ref{fig:selection} (left) is a plot of limiting mass versus 
redshift.  The uncertainties on the limiting mass correspond to the 
scatter about the best fit $\Delta\chi^{2}$-binding mass relation, 
determined using mock observations of the entire cluster ensemble at 
each redshift; this scatter is an indication of the 
morphological sensitivity of the detection threshold.  
We find that the limiting mass is independent of 
cosmological model-- at least for the four models tested.
Also in Figure \ref{fig:selection} (right) is a Press-Schechter estimate of the 
surface density of clusters more massive than 
$M_{lim}=2\times10^{14}h^{-1}_{50}$~$M_{\odot}$ as a function of redshift 
for three cosmological models 
(noted on the figure).  These models are all normalized to produce the 
correct cluster abundance at the present epoch.  Note the 
cosmological sensitivity of the cluster surface densities; we expect 
to find very few clusters with our survey if $\Omega=1$, whereas 
for more favored, low $\Omega_{0}$ models we expect hundreds of 
clusters \cite{holder99a}.

\begin{figure}[htb]
\vskip-10pt
\hbox to \hsize{
\includegraphics[width=0.55\textwidth]{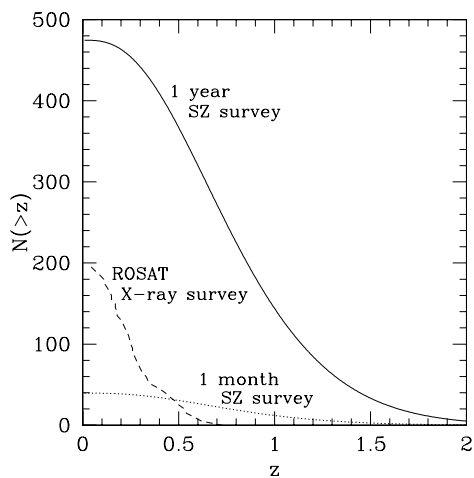}\hfill
\vbox to 2.6in{\hsize=2.0in
\caption[]{We plot the expected yield of our proposed SZE survey 
after one month (dotted) and one year (solid) if $\Omega_{0}=0.3$ 
and $\Omega_{\lambda}=0.7$.  For comparison we 
plot the yield of the ROSAT PSPC deep cluster survey (dashed).  After 
one month we will have more clusters at redshift $z>0.5$ than the entire 
ROSAT survey.  After one year we will have $\sim$100 clusters at 
redshifts $z>1$ and $\sim$400 clusters at redshifts $z.0.5$.
\label{fig:survey}}
\vfill}}
\vskip-10pt
\end{figure}

In Figure \ref{fig:survey} we plot the survey yield after one month 
(dotted line) and one year (solid line), where we have assumed that 
the currently favored model $\Omega_{m}=0.3$ and 
$\Omega_{\lambda}=0.7$ is correct (if $\Omega_{\lambda}=0$ the cluster 
count increases).  For comparison we also plot the yield of 200 
clusters from a deep ROSAT PSPC X-ray cluster survey \cite{vikhlinin98}.
Note that after a one month SZE survey we will have more redshift $z>0.5$ clusters 
than the deep ROSAT survey.  After one year we expect to have 
$\sim$100 clusters at redshifts $z>1$ and almost 400 clusters at 
redshifts $z>0.5$ \cite{holder99b}.  This highly sensitive probe for 
massive collapsed structures in the high redshift universe constitutes 
an exceedingly powerful test of current structure formation models.

The SZE survey itself will yield a list of galaxy clusters and 
constraints on the cluster surface density.  Cluster redshifts, 
masses, gas distributions and galaxy populations can then be examined
through followup observations with X-ray satellites and 8~m class 
telescopes like the VLT.  We anticipate that the scope and number of 
scientific issues addressed with followup 
observations of the high redshift clusters from the SZE survey 
will lead to enormous progress in cosmology.   

\medskip

\noindent JJM is supported by Chandra Fellowship grant PF8-1003, awarded
through the Chandra Science Center.  The Chandra Science Center is
operated by the Smithsonian Astrophyical Observatory for NASA under
contract NAS8-39073.  JEC acknowledges support from the David and Lucile 
Packard Foundation and the James S. McDonnell Foundation.

\clearpage
\addcontentsline{toc}{section}{Index}
\flushbottom
\printindex

\end{document}